\documentclass[conference]{IEEEtran}
\IEEEoverridecommandlockouts

\usepackage{cite}
\usepackage{amsmath,amssymb,amsfonts}
\usepackage{algorithmic}
\usepackage{graphicx}
\usepackage{multirow}
\usepackage{tabularx}
\usepackage[table]{xcolor} 
\definecolor{lightblue}{HTML}{B8F7F4}
\usepackage{textcomp}
\usepackage{xcolor}
\usepackage{url}
\def\BibTeX{{\rm B\kern-.05em{\sc i\kern-.025em b}\kern-.08em
    T\kern-.1667em\lower.7ex\hbox{E}\kern-.125emX}}
\begin{document}

\newcommand{\sthanks}[1]{\thanks{#1}}

\title{Bridge-SR: Schrödinger Bridge for Efficient SR}
\author{
    \IEEEauthorblockN{Chang Li*$^{2, 3}$\thanks{* Equally contributed. Work done during internship at Shengshu AI.}, Zehua Chen*$^{1, 2}$, Fan Bao$^{2}$, Jun Zhu$^{1, \dagger}$\thanks{$\dagger$ Corresponding author. e-mail: dcszj@tsinghua.edu.cn}}
    $\thanks{
    This work was supported by the NSFC Projects (Nos. 62350080, 92270001, 62106120, U24A20342), Tsinghua
    Institute for Guo Qiang, and the High Performance Computing Center, Tsinghua University. JZ was also supported
    by the XPlorer Prize.}$
    \IEEEauthorblockA{$^1$ Department of CST, Tsinghua University, Beijing, China}
    \IEEEauthorblockA{$^2$ Shengshu AI, Beijing, China}
    \IEEEauthorblockA{$^3$ University of Science \& Technology of China, Hefei, China}
}


\maketitle

\begin{abstract}
Speech super-resolution (SR), which generates a waveform at a higher sampling rate from its low-resolution version, is a long-standing critical task in speech restoration. Previous works have explored speech SR in different data spaces, but these methods either require additional compression networks or exhibit limited synthesis quality and inference speed.
Motivated by recent advances in probabilistic generative models, we present Bridge-SR, a novel and efficient any-to-48kHz SR system in the speech waveform domain. Using tractable Schrödinger Bridge models, we leverage the observed low-resolution waveform as a prior, which is intrinsically informative for the high-resolution target. By optimizing a lightweight network to learn the score functions from the prior to the target, we achieve efficient waveform SR through a data-to-data generation process that fully exploits the instructive content contained in the low-resolution observation.
Furthermore, we identify the importance of the noise schedule, data scaling, and auxiliary loss functions, which further improve the SR quality of bridge-based systems.
The experiments conducted on the benchmark dataset VCTK demonstrate the efficiency of our system: (1) in terms of sample quality, Bridge-SR outperforms several strong baseline methods under different SR settings, using a lightweight network backbone (1.7M); (2) in terms of inference speed, our 4-step synthesis achieves better performance than the 8-step conditional diffusion counterpart (LSD: 0.911 vs 0.927). Demo at https://bridge-sr.github.io.

\end{abstract}


\section{Introduction}
Speech signals often suffer from a limited sampling rate, which may result from hardware limitations, transmission issues, or inherent constraints in waveform generation models \cite{wang2023neural, liu2023audioldm, li2024quality}. In many application scenarios, this limited sampling rate restricts both the objective and perceptual quality of speech signals, necessitating an efficient super-resolution (SR) system to recover the high-frequency information.

In recent years, various speech SR systems have been proposed in the time domain~\cite{zhang2021wsrglow, lee2021nu, han2022nu, yu2023conditioning, zhu2024musichifi, koizumi2023miipher}, time-frequency domain~\cite{ku2024generative, shuai2023mdctgan, mandel2023aero, moliner2023solving}, and compressed space~\cite{liu2021voicefixer, liu2024audiosr, wang2023audit, kim2024audio, comunitaspecmaskgit}.
Among them, the methods in the time domain directly upsample the low-resolution waveform to its high-resolution version without relying on signal processing-based transformations (e.g., Short-Time Fourier Transform, STFT) or compression neural networks. This avoids cascading errors~\cite{liu2021voicefixer, liu2022neural}, adversarial training~\cite{kim2024audio, shuai2023mdctgan}, and large networks~\cite{zhang2021wsrglow, liu2024audiosr}.
In waveform space, diffusion models (DMs)-based SR systems \cite{ho2020denoising, kong2020diffwave} have achieved promising quality through their iterative refinement mechanism.
For example, NU-Wave \cite{lee2021nu} and NU-Wave2 \cite{han2022nu} condition DMs on the low-resolution waveform to generate the high-resolution target, achieving strong synthesis quality with a lightweight network. UDM+ \cite{yu2023conditioning} employs unconditional DMs and injects low-frequency constraints into the inference process.
However, these DMs-based SR systems rely on a \textit{noise-to-data} sampling trajectory, which is not straightforward when generating high-resolution waveforms from low-resolution observations. Moreover, it is challenging for generative models to efficiently synthesize high-resolution waveforms from uninformative Gaussian noise.
However, if we could exploit the strong prior information contained in the low-resolution waveform and leverage the advantages of the iterative refinement mechanism in sampling, there is still room for improvement in both synthesis quality and inference speed.

\begin{figure*}[htbp]
    \centering
    \includegraphics[width=0.912125\linewidth]{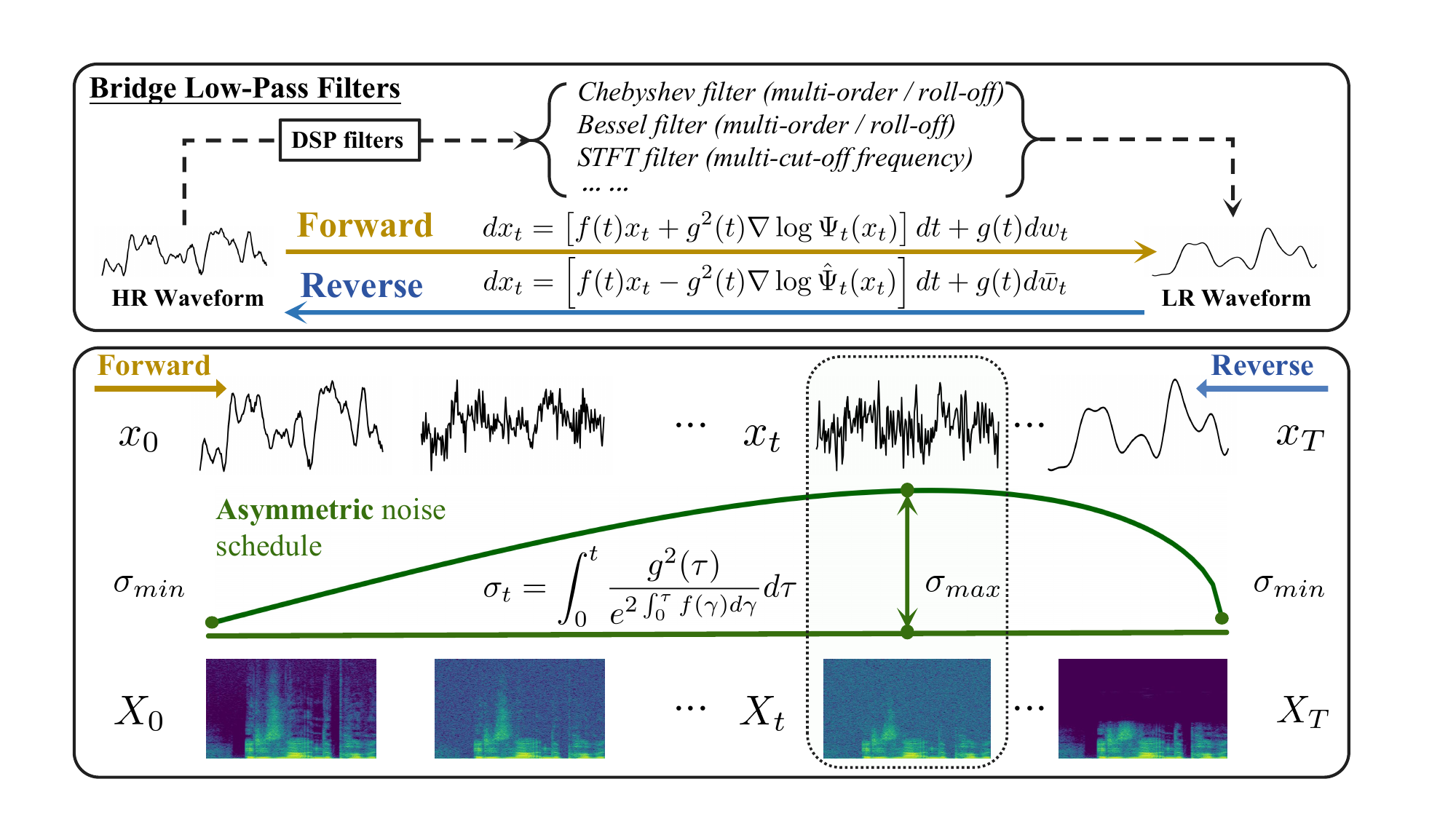}
    \caption{Overview of Bridge-SR. As shown in the upper part, the forward process of the Schrödinger bridge simulates low-pass filters (LPF). The lower part shows the intermediate representations in both the waveform and spectral domains under our asymmetric noise schedule.}
    \label{fig:main}
\end{figure*}

Inspired by recent progress in Schrödinger Bridges~\cite{chen2023schrodinger, liu20232}, we introduce Bridge-SR to fully utilize the informative low-resolution observation for high-resolution target generation. Specifically, we establish a tractable Schrödinger Bridge~\cite{chen2023schrodinger} between paired waveforms, using the observed low-resolution waveform as known prior distribution and generating the high-resolution target through a \textit{data-to-data} sampling trajectory.
By shifting from the \textit{noise-to-data} trajectory conditioned on the low-resolution waveform to a straightforward \textit{data-to-data} trajectory, our bridge models (BMs)-based SR system naturally exploits the instructive information in the prior, improving efficiency compared to DMs-based systems.


Furthermore, we investigate the noise schedules of BMs and propose a data scaling technique, which is important for capturing high-frequency information with small amplitudes. In addition, we introduce a fine-tuning process to incorporate auxiliary loss functions, further enhancing the performance of BMs on SR. In summary, we make the following contributions:

\begin{itemize} \item We make the first attempt to establish a Schrödinger bridge-based SR system that fully exploits the instructive information contained in low-resolution observations. \item We identify the importance of noise scheduling, data scaling, and auxiliary loss functions in SR, further improving the synthesis quality of BM-based SR systems. \item We conducted experiments on the VCTK~\cite{yamagishi2019cstr} benchmark dataset using a lightweight network with approximately 1.7M parameters~\cite{van2016wavenet, han2022nu}, achieving superior synthesis quality and faster inference speed for speech SR. \end{itemize}


\section{Bridge-SR}
As shown in Figure \ref{fig:main}, Bridge-SR utilizes a stochastic differential equation (SDE) based forward process to bridge the down-sampled waveform and its high-resolution version, and achieves the speech SR with the corresponding reverse-time SDE in sampling. The \textit{data-to-data} trajectory with investigated noise schedule helps us exploit the information contained in strong prior, realizing efficient SR.  

\subsection{Tractable Schrödinger bridge}

Given a low-resolution waveform $x_\text{LR}$, SR aims to reconstruct its high-resolution correspondence $x_\text{HR}$. During the training stage, $x_\text{LR}$ is derived from $x_\text{HR}$ through a series of signal processing filters with randomly selected settings. In score-based generative models (SGMs)~\cite{song2020score}, a forward SDE is defined between $x_0 = x_\text{HR} \sim p_\text{prior}$, and $x_T = x_\text{LR} \sim p_\text{data}$:
\begin{equation}
dx_t = f(x_t, t) \, dt + g(t) \, dw_t, 
\end{equation}
here, \( t \in [0, T] \) represents the current time step, with \( x_t \) denoting the state of data in the process. The drift and diffusion are given by the vector field \( f \) and the scalar function \( g \), \( w_t \) is the standard Wiener process, and the reference path measure \( p^\text{ref} \) describes the probability of paths from \( p_\text{prior} \) and \( p_\text{data} \).

Under the above framework with boundary constraints, the Schrödinger bridge (SB) problem~\cite{schrodinger1932theorie, chen2023schrodinger} seeks to find a path measure \( p \) of specified boundary distributions that minimizes the Kullback-Leibler divergence $D_\text{KL}$ between a path measure and reference path measure $p_\text{ref}$:
\begin{equation}
\min_{p \in \mathcal{P}_{[0,T]}} D_\text{KL}(p \parallel p^\text{ref})\quad s.t.\quad p_0 = p_\text{prior}, \;p_T = p_\text{data}
\label{eq:KL}
\end{equation}
where $\mathcal{P}_{[0,T]}$ denotes the collection of all path measures over the time interval ${[0,T]}$. 

In SB theory \cite{wang2021deep, chen2021likelihood, chen2023schrodinger}, this specific SB problem can be expressed as a pair of forward-backwords linear SDEs:
\begin{align}
d x_{t} &= \left[ f\left( x_{t}, t \right) + g^{2}(t) \nabla \log \Psi_{t} \left( x_{t} \right) \right] dt + g(t) dw_{t} \\
d x_{t} &= \left[ f\left( x_{t}, t \right) - g^{2}(t) \nabla \log \widehat{\Psi}_{t} \left( x_{t} \right) \right] dt + g(t) d\overline{w}_{t}
\end{align}
where the non-linear drifts \(\nabla \log \Psi_{t}(x_{t})\) and \(\nabla \log \widehat{\Psi}_{t}(x_{t})\) can be described by coupled partial differential equations (PDEs).
A closed-form solution for SB \cite{chen2023schrodinger} exists when Gaussian smoothing is applied with $p_{0} = \mathcal{N}\left(x_\text{HR}, \epsilon_{0}^{2} I\right)$ and $p_{T} = \mathcal{N}\left(x_\text{LR}, \epsilon_{T}^{2} I\right)$, to the original Dirac distribution. By defining $\alpha_{t} = e^{\int_{0}^{t} f(\tau) d\tau}$, $\bar{\alpha}_{t} = e^{\int_{1}^{t} f(\tau) d\tau}$, $\sigma_{t}^{2} = \int_{0}^{t} \frac{g^{2}(\tau)}{\alpha_{\tau}^{2}} d\tau$, and $\bar{\sigma}_{t}^{2} = \int_{t}^{1} \frac{g^{2}(\tau)}{\alpha_{\tau}^{2}} d\tau$, with boundary conditions and linear Gaussian assumption, tractable form of SB is solved as
\begin{equation}
\widehat{\Psi}_{t} = \mathcal{N}\left(\alpha_{t} x_\text{HR},\alpha_{t}^{2} \sigma_{t}^{2} I\right), \quad 
\Psi_{t} = \mathcal{N}\left(\bar{\alpha}_{t} x_\text{LR}, \alpha_{t}^{2} \bar{\sigma}_{t}^{2} I\right)
\end{equation} under $\epsilon_{T} = e^{\int_{0}^{T} f(\tau) d\tau} \epsilon_{0}$ and $\epsilon_{0} \rightarrow 0$ \cite{chen2023schrodinger}.
\begin{figure}[t]
    \centering
    \includegraphics[width=0.975\linewidth]{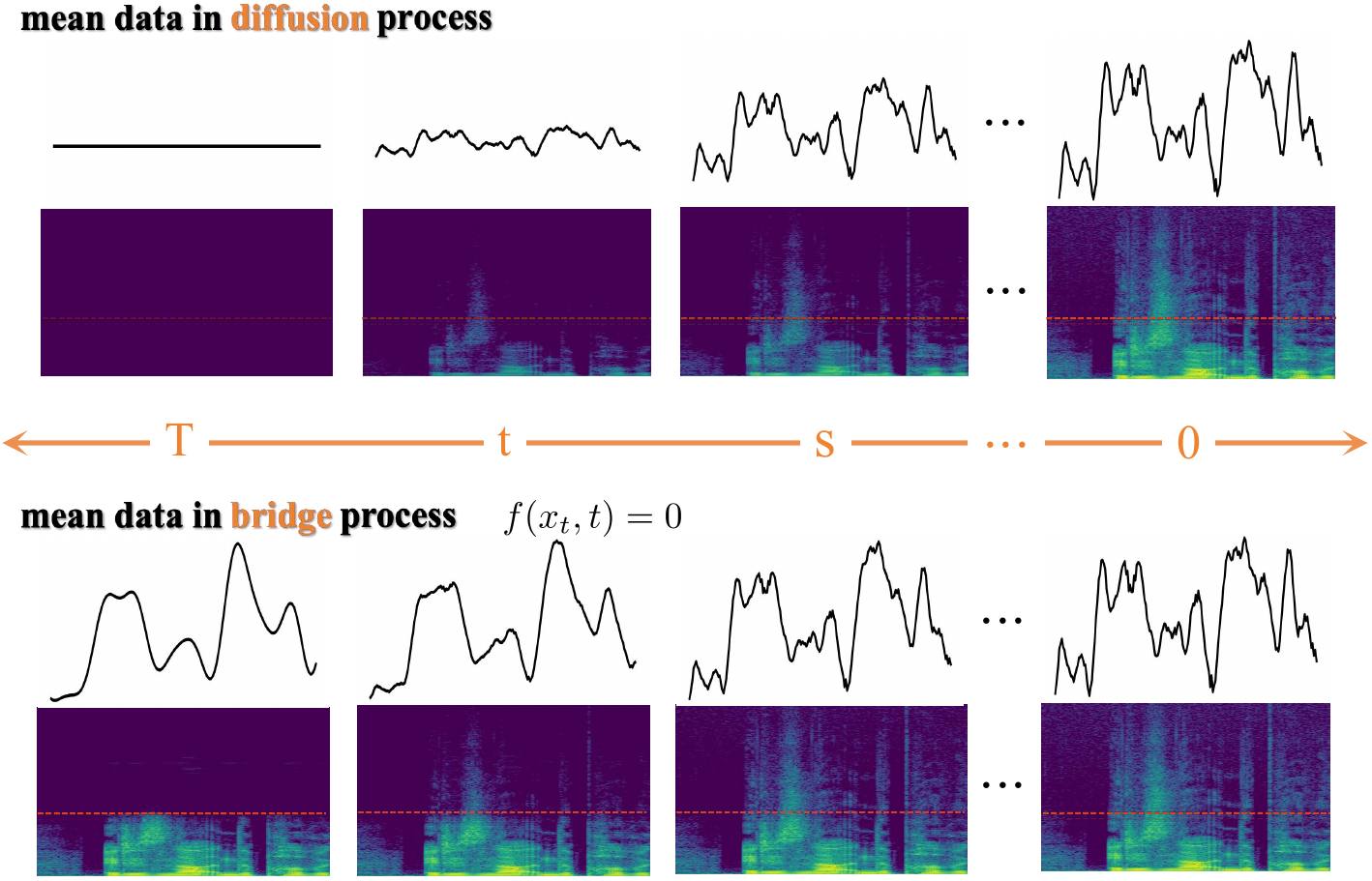}
    \caption{
    We show the means of the intermediate representations for the \textbf{diffusion process} and the \textbf{bridge process with no linear drift}, respectively. It becomes evident that, for the diffusion process, the low-frequency components gradually vanish during the forward SDE. In contrast, our bridge process preserves the low-frequency components.}
    \label{fig:data}
\end{figure}
Therefore, marginal distribution of $x_t$ at state $t$ is also Gaussian:
\begin{equation}
p_{t} = \Psi_{t}\widehat{\Psi}_{t} = \mathcal{N}\left(\frac{\alpha_{t}\bar{\sigma}_{t}^{2}}{\sigma_{1}^{2}} x_{0} + \frac{\bar{\alpha}_{t}\sigma_{t}^{2}}{\sigma_{1}^{2}}x_{T}, \frac{\alpha_{t}^{2}\bar{\sigma}_{t}^{2}\sigma_{t}^{2}}{\sigma_{1}^{2}} I\right).
\label{eq:p_t}
\end{equation}

During training, we calculate the bridge loss by directly predicting $x_0$ from randomly sampled $x_t$ from $p_t$:
\begin{equation}
\mathcal{L}_{\text{bridge}} = \mathbb{E}_{x_T \sim p_{\text{prior}}, x_0 \sim p_{\text{data}}} \mathbb{E}_{t} \left[ \left\| x_{\theta} \left(x_{t}, t, x_T\right) - x_{0} \right\|_{2}^{2} \right].
\label{eq:losss=}
\end{equation}

With uniformly sampled $t$, the filtering process is simulated by the forward SDE, while simultaneously optimizing a score estimator for the reverse solvers to execute SR.

\subsection{Noise scheduling and data scaling}
Bridge-TTS~\cite{chen2023schrodinger} and \cite{jukic2024schr} propose several noise schedules for the Schrödinger bridge, including Bridge-VP, Bridge-SVP, Bridge-$g_\text{max}$, and Bridge-$g_\text{const}$. In SR tasks, we empirically find that Bridge-$g_\text{max}$ outperforms Bridge-$g_\text{const}$, and both perform better than Bridge-(S)VP.
From the formulas for Bridge-$g_\text{max}$ and Bridge-$g_\text{const}$, which lack a linear drift (i.e., $f(t) = 0$), it follows that $\alpha(t) = 1$. Consequently, for any timestep $t \in [0, T]$, the data interpolation coefficient ${\alpha_t \bar{\sigma}_t^2 x_0} / {\sigma_1^2} + {\bar{\alpha}_t \sigma_t^2 x_T} / {\sigma_1^2}$ remains equal to 1. This indicates a constant low-frequency component throughout the SDE trajectory, as illustrated in Figure \ref{fig:data}. In contrast, Bridge-VP~\cite{chen2023schrodinger} and Bridge-SVP~\cite{jukic2024schr} offer variance-preserving processes during data trajectories, but they face similar issues as those in diffusion processes, where low-frequency constraints are not consistently maintained. This may impair the preservation of low frequencies and limit their interaction with high frequencies. As a result, fine-tuning VP processes for optimal SR outcomes becomes challenging.

For both Bridge-$g_\text{max}$ and Bridge-$g_\text{const}$, a linear schedule for \( g^2(t) = (1 - t)\beta_0 + t\beta_1 \) is employed, with the noise variance \( {\sigma_t^2 \bar{\sigma}_t^2} / {\sigma_1^2} \) reaching its peak at \( t = t_\text{p} \), where \( 2\sigma_{t_\text{p}}^2 = \sigma_1^2 \). When \( \beta_0 = \beta_1 \), a symmetric schedule with \( t_\text{p}/T = 1/2 \) is achieved, forming Bridge-$g_\text{const}$.
Although this symmetric schedule yields favorable results in tasks such as speech enhancement~\cite{wang2024diffusion} and image translation~\cite{liu20232}, we found that using an asymmetric schedule, where \(\beta_0 \rightarrow 0\) and \( t_\text{p}/T \approx 1/\sqrt{2} \), as in Bridge-$g_\text{max}$, allows the model to focus more on generating high-frequency components, leading to superior performance in the SR task compared to Bridge-$g_\text{const}$.

Furthermore, due to the inherently low energy of the high-frequency components, directly predicting $x_0 \sim p_\text{data}$ from $x_T \sim p_\text{prior}$ in the waveform space results in very low loss values during training, which hampers the effective optimization of the BMs to predict the data score function. 

To address this issue, we propose a scaling strategy to improve the model's ability to capture the high-frequency details. Specifically, we apply a scaling factor of $s = 1 / \sqrt{\text{Var}(x_\text{LR} - x_\text{HR})}$ to both $x_0$ and $x_T$, controlling the variance between the low-resolution and high-resolution waveforms to be as large as 1. This adjustment significantly enhances the model’s performance and training stability.
\label{B}

\subsection{Auxiliary losses}
In addition to optimizing \(\mathcal{L}_{\text{bridge}}\) to learn the high-resolution waveform's distribution, which has already yielded promising results. We found that incorporating an independent fine-tuning stage, which optimizes both the amplitude and phase of the STFT spectrum at each timestep, leads to further performance improvement. Specifically, we apply the perceptually weighted multi-scale STFT(mag) loss \(\mathcal{L}_\text{mag}\) following \cite{evans2024long}, and the multi-scale anti-wrapping phase loss \(\mathcal{L}_\text{phase}\) following \cite{ai2023neural} across \(M\) different resolutions between estimated $\hat{x_0} = x_\theta(x_t, t, x_T)$ and $x_0$:
\begin{equation}
    \mathcal{L}_{\text{aux}} = 
     \lambda_{\text{mag}} \sum_{r=1}^M \mathcal{L}_{\text{mag}}^{(r)} (\hat{x_0}, x_0)
    + \lambda_{\text{phase}} \sum_{r=1}^M \mathcal{L}_{\text{phase}}^{(r)} (\hat{x_0}, x_0).
\end{equation}
\label{C}
Incorporating the scaling factor in \ref{B}, the final loss function for fine-tuning can be summarized as $\mathcal{L}_{\text{final}} = \mathcal{L}_{\text{bridge}} + \mathcal{L}_{\text{aux}}$.

\begin{table*}[htbp]
\caption{Comparisons between Bridge-SR and baseline methods under different SR settings on VCTK test set. \\ SR denotes the sampling rate of the input waveform. The sampling rate of target audio is 48kHz.}
\label{table1}
\begin{center}
\setlength{\tabcolsep}{2pt} 
\begin{tabularx}{0.85\textwidth}{>{\centering\arraybackslash}p{1.6cm}| >{\centering\arraybackslash}p{0.7cm} | >{\centering\arraybackslash}X | >{\centering\arraybackslash}X | >{\centering\arraybackslash}X | >{\centering\arraybackslash}X | >{\centering\arraybackslash}X | >{\centering\arraybackslash}X | >{\centering\arraybackslash}X }

\hline
\textbf{Metrics} & \textbf{SR} & Input  & AudioSR~\cite{liu2024audiosr} & NVSR~\cite{liu2022neural} & mdctGAN~\cite{shuai2023mdctgan} & NU-Wave2~\cite{han2022nu} & UDM+~\cite{yu2023conditioning} & \textbf{Bridge-SR} \\
\hline\hline
\rowcolor{white}LSD $\downarrow$ & 24K & 2.997 & 0.876 & 0.845 & 0.809 & 0.740 & 0.769 & \textbf{0.716} \\
\rowcolor{white}LSD-LF $\downarrow$ & 24K & 0.201  & 0.482  & 0.441 & 0.397 & 0.423 & 0.219 & \textbf{0.202} \\
\rowcolor{white}LSD-HF $\downarrow$ & 24K & 4.234 & 1.132  & 1.104 & 1.070 & 1.011 & 1.064 & \textbf{0.992} \\
\rowcolor{white}SISNR $\uparrow$ & 24K & 31.23  & 23.76 & 22.14 & 28.61 & \textbf{30.21} & 24.63 & 29.17 \\
\hline
\rowcolor{white}LSD $\downarrow$ & 16K & 3.572  & 1.108 & 0.863 & 0.908 & 0.927 & 0.960 & \textbf{0.848} \\
\rowcolor{white}LSD-LF $\downarrow$ & 16K & 0.198  & 0.473  & 0.232 & 0.378 & 0.387 & 0.249 & \textbf{0.195} \\
\rowcolor{white}LSD-HF $\downarrow$ & 16K & 4.372  & 1.307 & 1.042 & 1.078 & 1.091 & 1.160 & \textbf{1.028} \\
\rowcolor{white}SISNR $\uparrow$ & 16K & 26.74  & 18.71 & 18.53 & 23.96 & 25.03 & 19.76 & \textbf{25.04} \\
\hline
\rowcolor{white}LSD $\downarrow$ & 12K & 3.835  & 1.177 & 0.972 & 1.006 & 1.015 & 1.081 & \textbf{0.928} \\
\rowcolor{white}LSD-LF $\downarrow$ & 12K & 0.203  & 0.465  & 0.433 & 0.390 & 0.349 & 0.229 & \textbf{0.191} \\
\rowcolor{white}LSD-HF $\downarrow$ & 12K & 4.428  & 1.327  & 1.091 & 1.138 & 1.148 & 1.240 & \textbf{1.065} \\
\rowcolor{white}SISNR $\uparrow$ & 12K & 23.41  & 15.70 & 15.49 & 20.63 & 22.16 & 17.68 & \textbf{22.48} \\
\hline
\rowcolor{white}LSD $\downarrow$ & 8K & 4.101  & 1.271& 1.018 & 1.051 & 1.140 & 1.251 & \textbf{1.015} \\
\rowcolor{white}LSD-LF $\downarrow$ & 8K & 0.188  & 0.383  & 0.370 & 0.344 & 0.290 & 0.217 & \textbf{0.184} \\
\rowcolor{white}LSD-HF $\downarrow$ & 8K & 4.492  & 1.379  & 1.102 & 1.139 & 1.234 & 1.367 & \textbf{1.101} \\
\rowcolor{white}SISNR $\uparrow$ & 8K & 20.05  & 12.97 & 12.97 & 18.41 & \textbf{19.30} & 14.78 & 19.02 \\
\hline
\rowcolor{white}Params $\downarrow$ & - & -  & 258.2M & 122.1M  & 103.0M*4 & \textbf{1.7M} & 2.3M & \textbf{1.7M} \\
\hline
\end{tabularx}
\end{center}
\end{table*}

\section{Experiments}
\subsection{Experimental Setup}
We conducted our experiments on the benchmark dataset for speech SR, VCTK \cite{yamagishi2019cstr}, where around 400 sentences are read by 108 English speakers, with each recording sampled at 48kHz.  
During training and inference, we mainly follow the settings of a strong diffusion baseline, NU-Wave2 \cite{han2022nu}, which is publicly available at \url{https://github.com/maum-ai/nuwave2}.  
For any-to-48kHz upsampling, the low-resolution input is uniformly sampled from 6kHz to 48kHz during training.  
We use a window length of 32768 (0.682 seconds at 48kHz), a batch size of 16, and a learning rate of \(5 \times 10^{-5}\).  
Bridge-SR is trained for 1M steps with a single bridge loss, and then fine-tuned for 70,000 steps with the auxiliary losses.  
The scaling factor is set to \(s = 12\), and the noise schedule is defined as \(g^2_\text{min} = 8 \times 10^{-7}\) and \(g^2_\text{max} = 8 \times 10^{-2}\).  
The weights of the auxiliary losses are set to \(\lambda_\text{mag} = 4 \times 10^{-6}\) and \(\lambda_\text{phase} = 5 \times 10^{-6}\).  
For evaluation, we test our system on SR tasks with 8kHz, 12kHz, 16kHz, and 24kHz input, upsampling to 48kHz waveforms.


\subsection{Baseline and Evaluation}
To provide a comprehensive evaluation, we compare Bridge-SR with 5 previous works, which include AudioSR~\cite{liu2024audiosr}, NVSR \cite{liu2022neural}, mdctGAN~\cite{shuai2023mdctgan}, NU-Wave2~\cite{han2022nu} and UDM+~\cite{yu2023conditioning}.
All models are trained with their official implementations~\cite{han2022nu} or directly tested with their publicly available checkpoints~\cite{liu2024audiosr, shuai2023mdctgan, yu2023conditioning}.
For evaluation, we follow previous works~\cite{liu2021voicefixer} to measure log-spectral distance (LSD) \cite{liu2021voicefixer} at full-band, low-frequency band (LSD-LF), and high-frequency band (LSD-HF) respectively. Furthermore, SI-SNR~\cite{liu2021voicefixer} is adopted to evaluate waveform-based synthesis quality.



\subsection{Inference Schedule}
\label{schedule_part}
Our diffusion counterpart, NU-Wave2, reports that high-quality SR is achieved with 8 sampling steps, and the synthesis quality does not improve further with an increase in the number of sampling steps~\cite{han2022nu}. In contrast, in Bridge-SR, we observe that a higher synthesis quality can be achieved with an increasing number of sampling steps.
In 50-step and 8-step sampling, we use a linear inference schedule between $t_\text{min}=10^{-5}$ and $t_\text{max}=1$ with the first-order PF-ODE sampler.
In few-step sampling, we test high-order samplers and use grid-searching algorithm for the choice of inference schedule.
In 4-step sampling, we employ the second-order SDE sampler with $t\in\{8 \times 10^{-2}, 5 \times 10^{-1}, 1\}$.
In 2-step and 1-step sampling, we employ a first-order PF-ODE sampler $t\in\{3 \times 10^{-2}, 9 \times 10^{-1}, 1\}$ and $t\in \{4 \times 10^{-2}, 1\}$ respectively.
The SDE and ODE samplers mentioned above are sourced from Bridge-TTS~\cite{chen2023schrodinger}.


\section{results}

\subsection{Results Analysis}
We show the comparison results of SR quality in Table~\ref{table1}.
As shown, with a lightweight network backbone (1.7M)~\cite{han2022nu}, Bridge-SR achieves the best quality in most evaluation metrics under different SR settings, outperforming the previous gan-based method~\cite{shuai2023mdctgan}, conditional diffusion models~\cite{han2022nu, liu2024audiosr}, and unconditional diffusion models~\cite{lee2021nu}. 
As the difference between Bridge-SR and NU-Wave2 mainly lies in the forward and reverse process, the significant improvement in SR quality achieved by Bridge-SR can be attributed to the data-to-data process realized by Schrödinger Bridge~\cite{chen2023schrodinger}, which is beneficial to fully exploit the instructive information provided by the low-resolution waveform.


\subsection{Ablation studies}
Table \ref{tab:my-table} shows our performance under different training noise schedules and sampling steps with 16kHz input, as well as the ablation studies for the scaling factor and the auxiliary loss proposed in Section~\ref{B} and Section~\ref{C}. 

\begin{table}[htbp]
\caption{Ablation studies under the setting of \textbf{16kHz to 48kHz} SR. \\ \textbf{SF} stands for the scaling factor, and we show  \textbf{the number of sampling steps} in parentheses.}
\label{tab:my-table}
\begin{center}
\setlength{\tabcolsep}{2pt} 
\begin{tabularx}{0.45\textwidth}{>{}p{1.80cm} >{\centering\arraybackslash}X >{\centering\arraybackslash}X >{\centering\arraybackslash}X >{\centering\arraybackslash}X >{\centering\arraybackslash}X }
\hline
              & LSD$\downarrow$   & L(LF)$\downarrow$  & L(HF)$\downarrow$ & SISNR$\uparrow$ & SSIM$\uparrow$~\cite{liu2021voicefixer}   \\ \hline\hline
NU-Wave2 & 0.927 & 0.387 & 1.091 & 25.03 & 0.769 \\  
\textbf{Ours $g_\text{max}$ (50)} & \textbf{0.848} & \textbf{0.195} & \textbf{1.028} & 25.04 & \textbf{0.800} \\  \hline
w $g_\text{const}$ (50) & 0.869 & 0.252 & 1.047 & 24.24 & 0.776 \\
w SVP (50) & 0.900 & 0.518 & 1.028 & 23.84 & 0.773 \\ 
w/o $\mathcal{L}_\text{aux}$ & 0.889 & 0.274 & 1.067 & 25.14 & 0.788  \\ 
w/o SF \& $\mathcal{L}_\text{aux}$ & 0.940 & 0.387 & 1.114 & 22.18 & 0.745 \\ \hline
Ours $g_\text{max}$ (8) & 0.913 & 0.212 & 1.107 & 23.93 & 0.787 \\ 
Ours $g_\text{max}$ (4) & 0.911 & 0.388 & 1.069 & \textbf{26.14} & 0.785 \\ 
Ours $g_\text{max}$ (2) & 0.947 & 0.560 & 1.063 & 24.27 & 0.784 \\ 
Ours $g_\text{max}$ (1) & 1.029 & 0.649 & 1.139 & 24.08 & 0.759 \\  
\hline
\end{tabularx}
\end{center}
\end{table}

The results demonstrate that both our scaling factor and auxiliary loss significantly improve SR performance. Among the noise schedules, the SVP schedule, which struggles to maintain low-frequency consistency during training, yielded the worst performance. In contrast, both the $g_\text{max}$ and $g_\text{const}$ schedules maintain low-frequency consistency. However, since the $g_\text{max}$ schedule allocates more sampling steps to generate high-frequency information compared to $g_\text{const}$, it achieves better results. This is consistent with the observations in the TTS synthesis \cite{chen2023schrodinger}. 
Furthermore, as the number of sampling steps decreases, 
Bridge-SR surpasses NU-Wave2 within 4 steps and remains competitive even with only 2 steps, further demonstrating the effectiveness of BMs for the SR task.

\section{Conclusion}

In this paper, we present Bridge-SR, establishing the first bridge-based SR system. By exploiting the instructive information contained in low-resolution waveform, we show improved synthesis quality and inference speed in comparison with DMs.

\bibliographystyle{IEEEtran}
\bibliography{ref}

\end{document}